\documentclass[shortnote]{jpsj3}

\title{Prerequisites for Relevant Spectral Density and Convergence of Reduced Density Matrices at Low Temperatures}

\author{Akihito Ishizaki}
\inst{Institute for Molecular Science, National Institutes of Natural Sciences and School of Physical Sciences, Graduate University for Advanced Studies, Okazaki, 444-8585, Japan}

\abst{Hierarchical equations of motion approach with the Drude-Lorentz spectral density has been widely employed in investigating quantum dissipative phenomena. However, it is often computationally costly for low-temperature systems because a number of Matsubara frequencies are involved. In this note, we examine a prerequisite required for spectral density, and demonstrate that relevant spectral density may significantly reduce the number of Matsubara terms to obtain convergent results for low temperatures.}

\begin{document}
\maketitle
\begin{figure}
\begin{center}
	\includegraphics{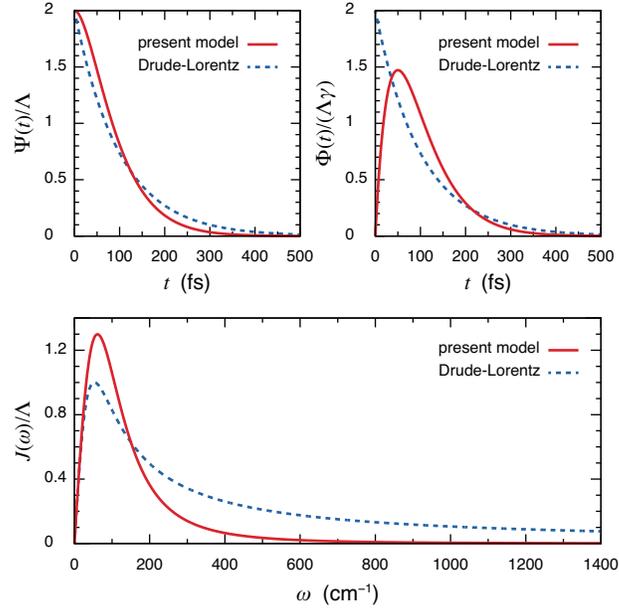}
\end{center}
	\caption{(Color online) 
	The relaxation function $\Psi(t)/\Lambda$, the response function $\Phi(t)/(\Lambda\gamma)$, and the spectral density $J(\omega)/\Lambda$ are presented for the employed model, eq.~\eqref{eq:relaxation-function} and the Drude-Lorentz model.  The bath characteristic time constant is set to $\tau_{\rm c} =100\,{\rm fs}$ ($\gamma = \Gamma/2 = 53\,{\rm cm^{-1}}$). The Matsubara frequency 	$\nu_1 = 2\pi/\beta\hbar$ take the value of $1310\,{\rm cm^{-1}}$ at temperature, 300\,K.}
	\label{fig:1}
\end{figure}

In condensed phase molecular systems, chemical dynamics are significantly impacted by the dynamics of the surrounding environment.\cite{Hanggi:1990km} 
A simple way to account for the environmental dynamics is to assume an exponential decay form for the correlation function of bath degrees of freedom, $\exp(-\gamma t)$.\cite{Kubo:1969ha} This corresponds to the Drude-Lorentz model and the spectral density is written as $J_{\rm DL}(\omega) = {2\Lambda\gamma\omega}/{(\omega^2+\gamma^2)}$, where $\Lambda$ quantifies the system-bath coupling strength. In quantum mechanical treatment, the correlation function is replaced with the symmetrized correlation function of the collective bath coordinate, $D(t)=(1/2)\langle \hat{X}(t) \hat{X}(0) + \hat{X}(0)\hat{X}(t) \rangle = (\hbar/\pi)\int^\infty_0 d \omega\,J(\omega)\coth(\beta\hbar\omega/2)\cos\omega t$, leading to
\begin{align}
	D_{\rm DL}(t)
	=
	2 \Lambda \tilde{E}_\beta(\gamma)  e^{-\gamma t}
	+
	\sum_{k=1}^\infty \frac{2 i}{\beta} J_{\rm DL}( i\nu_k) e^{-\nu_k t},
	\label{eq:DL-D-func}
\end{align}
where $\nu_k = 2\pi k/\beta\hbar$ is the bosonic Matsubara frequency and $\tilde{E}_\beta(\omega) = (\hbar\omega/2)\cot (\beta\hbar\omega/2) = \beta^{-1} + \beta^{-1} \sum_{k=1}^\infty {2\omega^2}/({\omega^2 - \nu_k^2})$ is introduced.
Also, the response function, $\Phi(t)=(i/\hbar)\langle [ \hat{X}(t), \hat{X}(0) ] \rangle = (2/\pi)\int^\infty_0 d \omega\,J(\omega)\sin\omega t$, is given by
\begin{align}
	\Phi_{\rm DL}(t)
	=
	2\Lambda\gamma e^{-\gamma t}.
	\label{eq:DL-Phi-func}
\end{align}
Note that transient behaviors of eqs.~\eqref{eq:DL-D-func} and \eqref{eq:DL-Phi-func} are coarse-grained.\cite{Kubo:1985bs}
Particularly, the response function in eq.~\eqref{eq:DL-Phi-func} gives a non-zero value at $t=0$ although it should vanish by definition, $\Phi(0) = (i/\hbar) \langle [ \hat{X}, \hat{X} ] \rangle= 0.$ 
Other problems arise owing to using the spectral density and the symmetrized correlation function. As shown in Fig.~\ref{fig:1}, the Drude-Lorentz spectral density exhibits a long tail in the high frequency region correspondingly to the coarse-grained nature. 
Consequently, a number of the Matsubara frequencies are required for the convergence of eq.~\eqref{eq:DL-D-func}. 
To compute quantum dynamics influenced by environments, the hierarchical equations of motion approach has been widely employed. \cite{Tanimura:1989bz,Ishizaki:2005jw,Xu:2005ka,Tanimura:2006ga}
However, the long-tail and convergence problems make the hierarchical equations approach for low temperature computationally costly. \cite{Ishizaki:2005jw} Hence, crude truncation was employed, \cite{Ishizaki:2009ky,Dijkstra:2012it} and a more sophisticated but applicable approach with Pad\'e approximant was proposed. \cite{Hu:2010ew,Hu:2011bw}

To address the the long-tail and convergence problems, we start with the relaxation function defined with the canonical correlation function, \cite{Kubo:1985bs} $\Psi(t)=\beta\langle \hat{X}(t);\hat{X}(0)\rangle$. The response function and spectral density are derived as  $\Phi(t) = - (d/d t) \Psi(t)$ and $J(\omega) = \omega \int^\infty_0 d t\, \Psi(t)\cos\omega t$, respectively. \cite{Kubo:1985bs} Hence, a prerequisite required for the relaxation function is
\begin{align}
	(d/d t) \Psi(t) \rvert_{t=0} = -\Phi(0) = 0.
	\label{eq:prerequisite}
\end{align}
As an example to satisfy this, we examine the relaxation function of the form,
\begin{align}
	\Psi(t) = 2\Lambda (1 + \Gamma t) e^{-\Gamma t}.
	\label{eq:relaxation-function}
\end{align}
In the limit of $\Gamma t \ll 1$, the function approaches the Gaussian form, $\Psi(t) \simeq 2\Lambda (1 - \Gamma^2 t^2) \simeq 2\Lambda e^{-\Gamma^2 t^2}$, providing the relevant short-time approximation. \cite{Kubo:1985bs} The characteristic time constant is obtained as $\tau_{\rm c} = \int^\infty_0 d t\, \Psi(t)/\Psi(0) = 2/\Gamma$, and thus, $\Gamma/2$ corresponds to $\gamma$ in the Drude-Lorentz model. The response function and spectral density are obtained as $\Phi(t) = 2\Lambda\Gamma^2 t e^{-\Gamma t}$ and $J(\omega) = {4\Lambda \Gamma^3 \omega}/{(\omega^2+\Gamma^2)^2}$, respectively. 
In the limit of $\omega \ll \Gamma$, the spectral density is written as the Ohmic form, $J(\omega) = 4\Lambda\Gamma^{-1} \omega$.
As shown in Fig.~\ref{fig:1}, this spectral density does not exhibit a long tail, indicating that the tail of the Drude-Lorentz spectral density is governed by the condition in eq.~\eqref{eq:prerequisite}. Therefore, it is expected that the symmetrized correlation function associated to eq.~\eqref{eq:relaxation-function} converges with a smaller number of the Matsubara frequencies.

Note that the time-varying characteristics of all the two-body correlation functions need to take exponential decay forms to perform the hierarchical expansion for the equation of motion. Hence, we rewrite eq.~\eqref{eq:relaxation-function} into the form, $\Psi(t) = 2\Lambda ( \cos\epsilon t + \Gamma \epsilon^{-1} \sin \epsilon t ) e^{-\Gamma t}$ with $\epsilon$ denoting an infinitesimal number, $\epsilon \to 0$. Practically, the value of $\epsilon$ is chosen as $\lvert \epsilon \rvert \lesssim \Gamma/4$, modifying the characteristic time constant as $\tau_{\rm c} = [(\Gamma^2+\epsilon^2)/2\Gamma]^{-1}$. This form is a Brownian oscillator, \cite{Mukamel:1995us,Tanaka:2009fn} and the response function and spectral density are given by $\Phi(t) = 2\mathrm{Re}[ i \epsilon^{-1} \Lambda \Gamma_+\Gamma_- e^{-\Gamma_+t} ]$ and $J(\omega) = 4\Lambda \Gamma \Gamma_+ \Gamma_- \omega/[(\omega^2+\Gamma_+^2)(\omega^2+\Gamma_-^2)]$, respectively, where $\Gamma_\pm = \Gamma \pm i\epsilon$ is introduced. 
The symmetrized correlation function is calculated as $D(t) = D_{\rm cl}(t) + \sum_{k=1}^\infty D^{(k)}(t)$, where $D_{\rm cl}(t) = \beta^{-1} \int^\infty_t d s\, \Phi(s)$ is the classical correlation function, $D_{\rm cl}(t) = 2\mathrm{Re}[i (\beta\epsilon)^{-1} \Lambda \Gamma_- e^{-\Gamma_+ t}]$, and $D^{(k)}(t)$ is the $k$-th quantum correction term defined by
\begin{align}
	D^{(k)}(t) 
	&=
	2\mathrm{Re}
	\left[
		\frac{i\Lambda \Gamma_-}{\epsilon} 
		\tilde{E}^{(k)}_\beta(\Gamma_+)
		e^{-\Gamma_+ t}
	\right]
	+
	\frac{2i}{\beta} J(i\nu_k) e^{-\nu_k t}
	\label{eq:quantum-correction}
\end{align}
with $\tilde{E}^{(k)}_\beta(\omega) = \beta^{-1}{2\omega^2}/({\omega^2 - \nu_k^2})$.

For numerical demonstration of the hierarchical equations for eq.~\eqref{eq:relaxation-function}, we consider two typical examples of quantum dissipative processes in molecular systems: longitudinal relaxation between energy levels and outer-sphere (or Marcus-type) electron transfer reaction. 
In either model the system-plus-bath Hamiltonian is written as $\hat{H}_{\rm tot} = \hat{H}_{\rm sys} + \hat{H}_{\rm bath} + \hat{H}_{\rm int}$.
However, individual models require different forms of the system-bath interaction Hamiltonian $\hat{H}_{\rm int}$.

\begin{table}
	\begin{tabular}{r|cccc}
	\hline	
	temperature		&	classical correlation	&	1st-order corr.	&	2nd-order corr.
	\\
	\hline
		$300\,{\rm K}$	&	0.2947		&	0.3020	&	0.3020
	\\
		$150\,{\rm K}$	&	0.1396		&	0.1790	&	0.1790
	\\
		$100\,{\rm K}$	&	0.0150		&	0.1019	&	0.1018
	\\
		$75\,{\rm K}$	&	$-0.0897$	&	0.0473	& 	0.0472
	\\
		$60\,{\rm K}$	&	$-0.1808$	&	0.0026	&	0.0025
	\\
	\hline
	\end{tabular}
	\caption{
	Density matrix element $\langle 1 \vert \hat\rho \vert 1 \rangle$ at the equilibrium state at varying temperatures for the spin-boson Hamiltonian, eq.~\eqref{eq:spin-boson-hamiltonian}. The energy gap is fixed at $E_1 - E_2 = 200\,{\rm cm^{-1}}$. The bath characteristic time constant is set to be $\tau_{\rm c} =100\,{\rm fs}$ ($\Gamma = 100\,{\rm cm^{-1}}$ and $\epsilon= 25\,{\rm cm^{-1}}$), and the system-bath coupling strength is $\Lambda = 50\,{\rm cm^{-1}}$. Numerical calculations were done with the hierarchical equations of motion for the bath classical correlation function, and the first- and second-order low-temperature corrections.
	}
	\label{table:1}
\end{table}

The longitudinal relaxation can be modeled by a simple spin-boson Hamiltonian,  
\begin{gather}
	\hat{H}_{\rm sys} = E_1 \lvert 1 \rangle\langle 1 \rvert + E_2 \lvert 2 \rangle\langle 2 \rvert,
	\label{eq:spin-boson-hamiltonian}
\\
	\hat{H}_{\rm int} = \hat{X}(\lvert 1 \rangle\langle 2 \rvert + \lvert 2 \rangle\langle 1 \rvert).
\end{gather}
The relaxation is described as the one-phonon process and hence the energy gap $\lvert E_1 - E_2 \rvert$ needs to resonate with spectral density $J(\omega)$. Table~\ref{table:1} presents the density matrix element $\langle 1 \vert \hat\rho \vert 1 \rangle$ at the equilibrium state at several temperatures for the fixed energy gap, $E_1 - E_2 = 200\,{\rm cm^{-1}}$. 
The bath parameters are set to $\Lambda = 50\,{\rm cm^{-1}}$, $\Gamma = 100\,{\rm cm^{-1}}$ and $\epsilon= 25\,{\rm cm^{-1}}$, which correspond to the spectral density in Fig.~\ref{fig:1}. Hence, the energy gap resonates with the spectral density.
At lower temperatures, the Matsubara frequency $\nu_k$ is smaller, and thus the contribution of $i J(i\nu_k)$ in eq.~\eqref{eq:quantum-correction} increases. When the low-temperature corrections are not employed, the density matrix elements for $T = 75\,{\rm K}$ and $T = 60\,{\rm K}$ become negative. However, the inclusion of the first-order correction term gives convergent results.

\begin{table}
	\begin{tabular}{r|ccc}
	\hline	
	$E_1^\circ - E_2^\circ$		&	classical correlation	&	canonical distribution
	\\
	\hline
		$200\,{\rm cm^{-1}}$	&	0.2756			&	0.2770
	\\
		$400\,{\rm cm^{-1}}$	&	0.1267			&	0.1280
	\\
		$600\,{\rm cm^{-1}}$	&	0.0527			&	0.0532
	\\
		$800\,{\rm cm^{-1}}$	&	0.0211			&	0.0211
	\\
		$1000\,{\rm cm^{-1}}$	&	0.0085			&	0.0081
	\\
	\hline
	\end{tabular}
	\caption{Density matrix element $\langle 1 \vert \hat\rho \vert 1 \rangle$ at the equilibrium state at temperature $300\,{\rm K}$ for varying values of $E_1^\circ - E_2^\circ$ in the electron transfer Hamiltonian, eq.~\eqref{eq:ET-hamiltonian}. The inter-site coupling, the reorganization energy, and the bath characteristic time are $J_{12}=50\,{\rm cm^{-1}}$, $\Lambda = 500\,{\rm cm^{-1}}$, and $\tau_{\rm c} =100\,{\rm fs}$ (with $\Gamma = 100\,{\rm cm^{-1}}$ and $\epsilon= 25\,{\rm cm^{-1}}$), respectively. Numerical calculations were done with the hierarchical equations of motion for the classical correlation function. Values of the canonical distribution are also shown.
	}
	\label{table:2}
\end{table}

The outer-sphere electron transfer reaction is described with the Hamiltonian,\cite{Fujihashi:2018hb}
\begin{gather}
	\hat{H}_{\rm sys} 
	= 
	E_1^\circ \lvert 1 \rangle\langle 1 \rvert + (E_2^\circ + \Lambda) \lvert 2 \rangle\langle 2 \rvert + V_{12} (\lvert 1 \rangle\langle 2 \rvert + {\rm h.c.}),
	\label{eq:ET-hamiltonian}
	\\
	\hat{H}_{\rm int} = \hat{X} \lvert {\rm 2} \rangle\langle {\rm 2} \rvert
\end{gather}
where $E_{\rm 1/2}^\circ$ is the equilibrium state energy of the donor/accepter state.
The system-bath coupling strength $\Lambda$ is the reorganization energy for the electron transfer, and the collective bath coordinate $\hat{X}$ corresponds to the solvation coordinate. 
The electron transfer reaction is a thermally activated barrier crossing process; thus, resonance between the energy gap $\lvert E_1^\circ - E_2^\circ \rvert$ and the spectral density $J(\omega)$ is unnecessary. 
Table~\ref{table:2} presents the density matrix element $\langle 1 \vert \hat\rho \vert 1 \rangle$ at the equilibrium state at temperature $300\,{\rm K}$ for varying values of the energy gap, $E_1^\circ - E_2^\circ$. The inter-site coupling and the reorganization energy are $J_{12}=50\,{\rm cm^{-1}}$ and $\Lambda = 500\,{\rm cm^{-1}}$, respectively.\cite{Fujihashi:2018hb} The bath characteristic time (timescale of the solvation dynamics) is set to $\tau_{\rm c} =100\,{\rm fs}$ with $\Gamma = 100\,{\rm cm^{-1}}$ and $\epsilon= 25\,{\rm cm^{-1}}$, correspondingly to the spectral density in Fig.~\ref{fig:1}.
 As presented in Table~\ref{table:2}, practically convergent equilibrium populations are obtained at low temperatures of $\beta\hbar(E_1^\circ-E_2^\circ) > 1$ even in the absence of the low-temperature correction terms. This is explained as follows:
At the fixed temperature 300\,K, the Matsubara frequency $\nu_k$ takes the value of $1310\,{\rm cm^{-1}} \times k$. Hence, $i J(i \nu_k)$ in eq.~\eqref{eq:quantum-correction} is insignificant when the spectral density $J(\omega)$ presented in Fig.~\ref{fig:1} is considered.

In summary, we discussed the the prerequisite required for relevant spectral density or equivalently relaxation functions, eq.~\eqref{eq:prerequisite}. The spectral density that satisfies the prerequisite does not exhibit a long tail in the high frequency region. Consequently, a smaller number of the Matsubara frequencies are required for the convergence in the symmetrized correlation function, potentially reducing the computational cost to obtain convergent results in the hierarchical equations approach for low temperature systems.

\begin{acknowledgment}
The author thanks Akihito Kato, Yuta Fujihashi, and Tatsushi Ikeda for their critical reading. This work was supported by JSPS KAKENHI Grant No.~17H02946 and MEXT KAKENHI Grant No.~17H06437 in Innovative Areas ``Innovations for Light-Energy Conversion.''
\end{acknowledgment}


\end{document}